\documentclass[%
 reprint,
superscriptaddress,
frontmatterverbose, 
showpacs,preprintnumbers,
 amsmath,amssymb,
 aps,
pra,
]{revtex4-1}
\usepackage{subfigure}
\usepackage{xcolor}
\usepackage{graphicx}
\usepackage{dcolumn}
\usepackage{bm}

\begin{document}


\title{Intensity and radiation statistics of correlated dipoles in GHZ and W-states}

\author{Shaik Ahmed }
 \affiliation{School of Physics, University of Hyderabad, Hyderabad - 500046, India}
\author{Prasanta K. Panigrahi}%
\affiliation{Indian Institute of Science Education and Research(IISER), Kolkata - 700106, India}

\author{P. Anantha Lakshmi$^1$}
%
\email{pochincherla03@gmail.com,  palsp@uohyd.ernet.in}


\date{\today}

\begin{abstract}
We investigate the super and sub-radiance characteristics of the radiation emitted from a system of three two-level atoms, in the GHZ and W-states.  The dipolar coupling between atoms leads to two distinct configurations, one in which the three atoms are on a line, and the other in a closed loop form, wherein each atom interacts with both its neighbours.  The quantum statistical properties of the emitted radiation show distinctly different characteristics for the GHZ and W-states resulting from the above configurations.  The far-field radiation pattern also shows distinct differences between differently entangled states, making it possible for optical probing of entangled states. 

\begin{description}
\item[keywords]
 Two-level atoms, W-states, GHZ states, photon statistics, Entanglement     
\end{description}
\end{abstract}
\maketitle

\section{Introduction} 
Superradiance, first introduced by Dicke\cite{r1}, is the coherent spontaneous emission generated due to the cooperative effects among atoms \cite{r22}.  It has been extensively studied in the literature, with the identification of a phase transition, separating the coherent phase of radiation from the incoherent one \cite{r2,r3,r4}.  It has attracted significant interest due to its wide range of possible applications, ranging from generation of X-ray lasers with high powers\cite{r5}, short pulse generation\cite{r6} and self-phasing in a system of classical oscillators \cite{r7}, to name a few.  
Superradiance has been studied experimentally in many physical systems \cite{r9,r10,cs}.  It can provide insights into various applications of the entangled atomic ensembles and generated quantum states of light for quantum memories \cite{bc,rr}, quantum communication \cite{mde}, quantum cryptography \cite{aks,ak} and quantum information\cite{dp}.  
In the context of quantum information, it is of particular interest to explore how the super radiant behavior gets affected for a collection of atoms, when the states are entangled in different ways.  This will allow optical probe of entangled states and may throw light on the nature of entanglement.  It is worth mentioning that some of the excited and ground states in the original study of Dicke are highly entangled. 
Depending on the nature of interactions on a multi-particle system, one can realise different types of entangled states\cite{gh}.  As is well-known, Bell states \cite{nc} are the maximally entangled states for two particles.  

The entanglement characteristics of three or more particle systems is yet to be completely understood.  For three particles, from the perspective of teleportation,  the states can be classified into two categories, the GHZ \cite{r12} and the W-classes\cite{r11}.    The latter can be symmetric or asymmetric type, having different entanglement properties.    It has been shown that symmetric W-state fails to teleport although it can carry out other quantum tasks\cite{r11}.  These states have found lot of practical importance in the field of quantum information processing\cite{r13}.  Schemes using some of these states \cite{mb,HH,AR,SZ,CSY} have been proposed for perfect teleportation \cite{r14,r15} as well as for super dense coding\cite{r16}.  A variety of experimental schemes for generation of the three - qubit GHZ and W - states involving the optical methods \cite{Eibl,Bouw}, cavity QED \cite{Haroche} and  ion trap techniques \cite{CFR} have been reported.  In particular, an NMR scheme to generate these W - state as well as GHZ - states, involving the unitary evolution of the XY Hamiltonian followed by rotation of the appropriate qubits, is proposed \cite{rkr-QI}.

Recently, Wiegner et al.\cite{r17} have investigated the super and sub-radiant characteristics of an N-atom system prepared in generalized W-states of the  form $\dfrac{1}{\sqrt{n}}|n-1,1> $, with one atom in the ground state and (n - 1) atoms  in the excited states.  These states can be generated from any N-qubit interactions, which conserve the $ \widehat{z} $ component of the total spin of the N - qubits.  Deng et al.\cite{r18} have proposed several schemes for generation of these states.  The entanglement of the W- state is robust, as compared to the GHZ state against particle loss.  It remains entangled even after any n-2 parties lose information about their particles \cite{r11}.

   Here we examine a system of three two-level atoms coupled by dipolar interactions and investigate a regime where the inter atomic distance is smaller than the emission wavelength.  Two possible configurations of the chain of three atoms, open and closed, are studied systematically.  For each of these configurations, one can generate two types of W - states, namely one atom in the excited state and two atoms in the excited state and vice-versa.  In certain parameter regime, GHZ-states are also  generated.  It is expected that the dipolar field configuration as well as the radiation characteristics will carry signature of entanglement.  The specific superpositions underlying these states will influence the radiation pattern.  In the following, we carry out a systematic investigation of the field configuration on the far zone and the super and sub-radiation properties of the emitted radiation. 
   
 The paper is organized as follows.
 In sec.II, we give the Hamiltonian for the system of three identical two-level atoms interacting with each other via dipole-dipole coupling. In Sec.III, we present exact results for the intensity of the emitted radiation from the three two-level atoms arranged in line configuration.  We show how the nature of initial entangled W - state influences its radiative characteristics leading to superradiant/sub-radiant emission of photons.  This is followed with the results of the intensity, for the case, where two atoms are in the excited state.  In Sec. IV, we present the corresponding results for the atoms arranged in a loop - configuration, with exact expressions for the intensity-intensity correlation, as well as far-field pattern.  Conclusions drawn from this study are presented in Section V, wherein we provide direction for further investigation.
 
 \section{The Model} 
The Hamiltonian for the system of three identical two-level atoms coupled through dipole-dipole interaction is given by
 \begin{equation}\label{eq:Hamiltonian}
H={\omega \sum^{3}_{i=1}}S^{z}_{i}+\sum^{3}_{i\neq j=1}\Omega_{ij}(S^{+}_{i}S^{-}_{j}+H.C), 
\end{equation} 

The first term describes the unperturbed energy of the system with the second one representing the dipole - dipole interaction  between the atoms, where  $ \Omega_{ij} $, the dipole dipole interaction strength, is a function of the inter-atomic separation `$ d $'. 
In the above, $ \omega $ is the atomic transition frequency,  $ S^{+}_{i} = |1\rangle_{i} \langle 0| $  and $S^{-}_{i}=|0 \rangle_{i} \langle 1| $  are the raising and lowering operators of the $i^{th}$ atom. 
\begin{figure}[h!]
\centering
\includegraphics[width=4 cm]{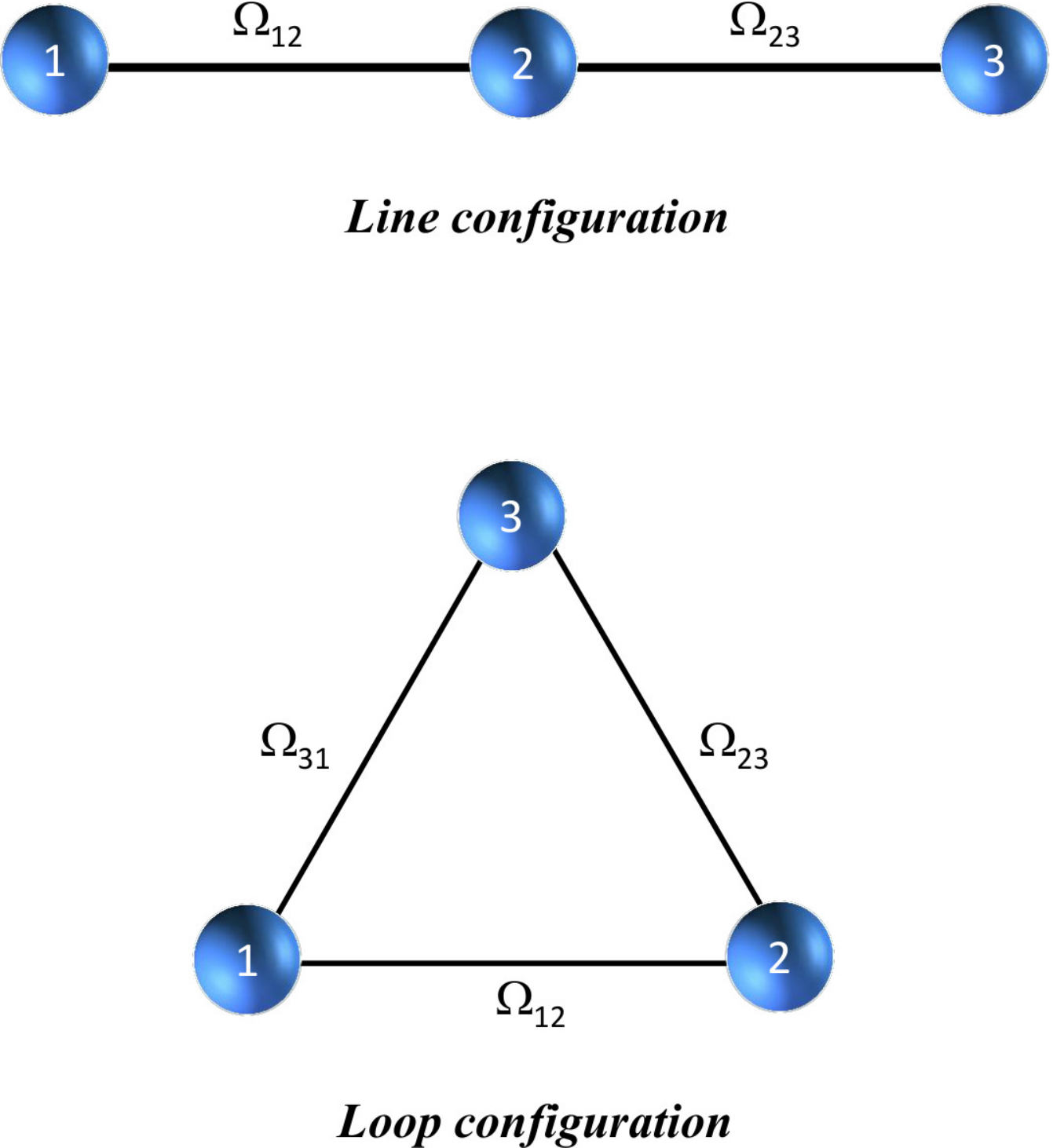}
\caption[]{(Color online) Schematic representation of Line(open loop) and Loop (closed loop) configurations for three qubits coupled via dipole-dipole interaction.}
\label{fig:One}
\end{figure}
  
Linear and closed loop configurations are the two distinct ones, in which three atoms can be arranged.  In the former configuration, the atoms are located in a linear array, wherein the interactions
 between successive atoms are considered: the interaction between atoms 1 and 2, and that between atoms 2 and 3 alone are taken into account.  The second configuration considered is a closed loop, where each of the three atoms interacts with its neighbours.  Both these configurations are shown in figure \ref{fig:One}.  Any other arrangement can be seen to be a simple variation of these two units.  
 As the atom-atom couplings are in different in both these configurations, the results for field intensities and radiation statistics for the two configurations are found to be markedly different.  
 In the absence of any radiation field, the Hamiltonian can be expanded in the standard basis, $ | ijk \rangle $, with $ i,j,k=0,1$.  
 The presence of dipole coupling between the atoms causes a mixing of the energy levels leading to creation of entangled states.  Depending on the number of atoms that are in the excited state, two different type of entangled states manifest.  In the first case, only one atom is in the excited state and in the second, two atoms are in the excited state generating different types of W-states.  The GHZ state is also generalized in both the configuration.
 
\section{The intensity characteristics of light emitted by three atoms in a Line-configuration}
\begin{figure}[ht!]
\centerline{\includegraphics[width=6.25 cm,height=6.0 cm]{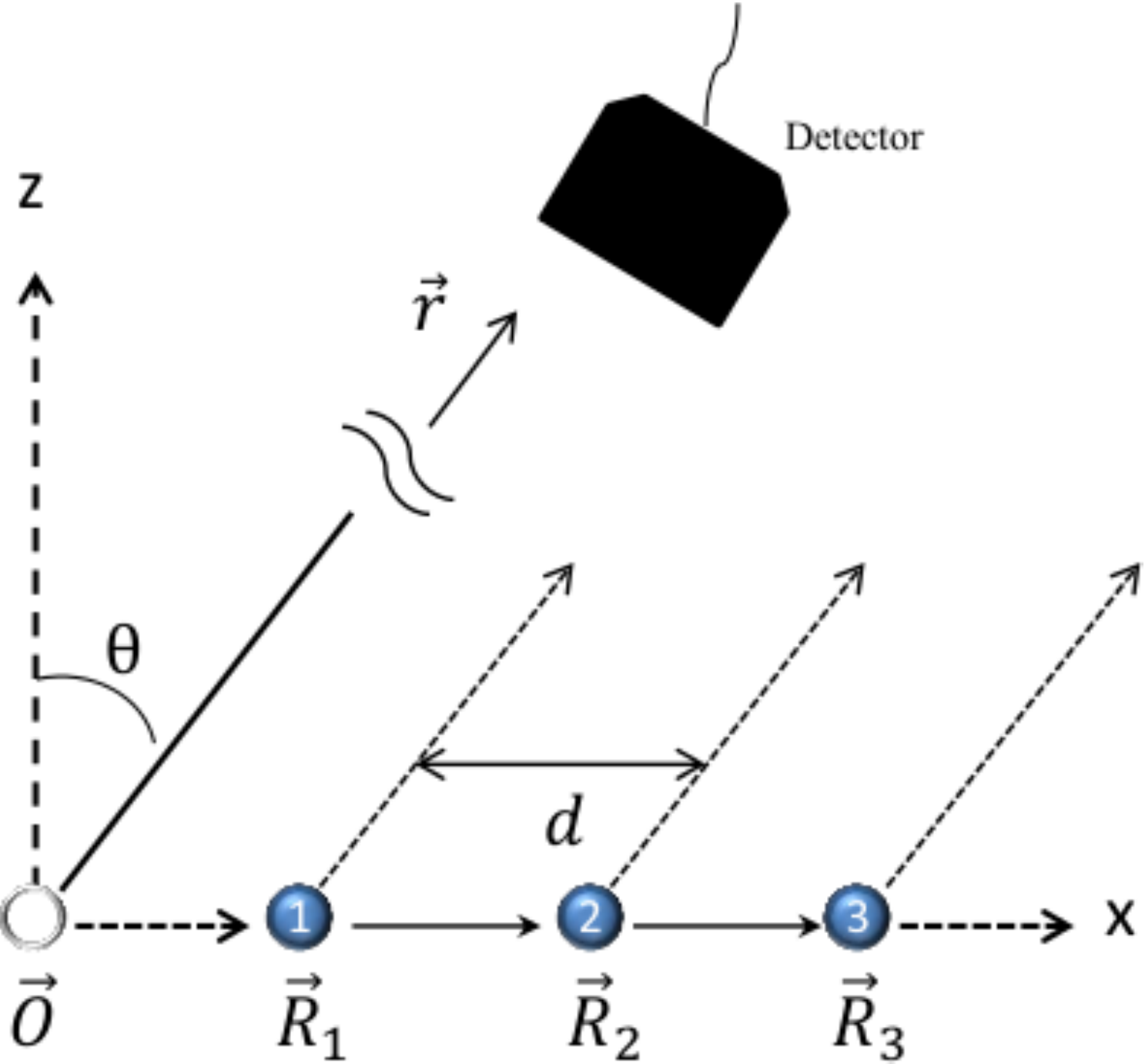}}
\caption[]{(Color online) Schematic diagram of the system: where three identical two-level atoms are localized at positions $ \bar{R}_{1}$ to $\bar{R}_{3}$.  A detector at position $ \bar{r} $ records a photon scattered by the atoms, in the far field regime.}
\label{fig:Two}
\end{figure}

As mentioned earlier, in the line configuration, we consider a system of three identical dipole coupled two-level atoms placed symmetrically along a line.
 These atoms are localized at positions $ R_{1},R_{2} $ to $ R_{3} $, where for simplicity, we consider equal spacing $ d $ between adjacent atoms, depicted in Fig. \ref{fig:Two}, with $\Omega_{12}=\Omega_{23}>>\Omega_{31}$.

The intensity emitted by this three atom system in the far field zone is investigated.  For this purpose, the detector is placed at position $ \vec{r} $ and the positive frequency component of the electric-field operator \cite{gsa} is considered,
 \begin{equation}\label{eq:E}
\hat{E}^{(+)}=-\frac{e^{ikr}}{r}\sum_{j}\vec{n}\times(\vec{n}\times\vec{p}_{ge})e^{-i\phi_{j}}\hat{S}^{-}_{j}
\end{equation}
where $\vec{r}$ indicates the detector position,
the unit vector $ \vec{n}=\frac{\vec{r}}{r} $ and $ \vec{p}_{ge} $ the dipole moment of the transition $ |e \rangle \rightarrow |g \rangle $. Here  $ \phi_{j} $ is the relative optical phase accumulated by a photon emitted at $ \vec{R}_{j} $ and detected at $ \vec{r} $, given by 
\begin{equation}
\phi_{j}(\vec{r})\equiv \phi_{j}=k\vec{n}.\vec{R}_{j}=jkd\sin{\theta}.
\end{equation} 

We also assume $ \vec{p}_{ge} $ to be along the y direction and $ \vec{n} $ in the x-z plane, so that $ \vec{p}_{ge}.\vec{n} =0$.  These assumptions give rise to dimensionless expressions for the amplitudes and hence intensities, resulting  in the following expression for the radiated intensity at $ \vec{r} $
 \begin{flalign*}
I(\vec{r})&=\sum_{i,j}\langle \hat{S}^{+}_{i}\hat{S}^{-}_{j}\rangle e^{i(\phi_{i}-\phi_{j})}&
\end{flalign*}
{\begin{scriptsize}\begin{eqnarray}
&=&\sum_{i}\langle \hat{S}^{+}_{i}\hat{S}^{-}_{i} \rangle +
\left(\sum_{i\ne j}\langle \hat{S}^{+}_{i}\rangle \langle \hat{S}^{-}_{j}\rangle +
\sum_{i\ne j}(\langle \hat{S}^{+}_{i}\hat{S}^{-}_{j}\rangle -\langle \hat{S}^{+}_{i}\rangle \langle \hat{S}^{-}_{j}\rangle)\right )e^{i(\phi_{i}-\phi_{j})} \nonumber\\
\end{eqnarray}\end{scriptsize}}

Thus, the characteristics of the intensity would depend on the incoherent terms $ \langle \hat{S}^{+}_{i}\hat{S}^{-}_{i}\rangle $, the non vanishing of the dipole moments $\langle \hat{S}^{+}_{i}\rangle$ and the quantum correlations like $ \langle \hat{S}^{+}_{i}\hat{S}^{-}_{j}\rangle - \langle \hat{S}^{+}_{i}\rangle \langle \hat{S}^{-}_{j}\rangle $.

\begin{figure}[ht!]
\centering
\subfigure[]{
\includegraphics[width=7.5 cm,height=5.0 cm]{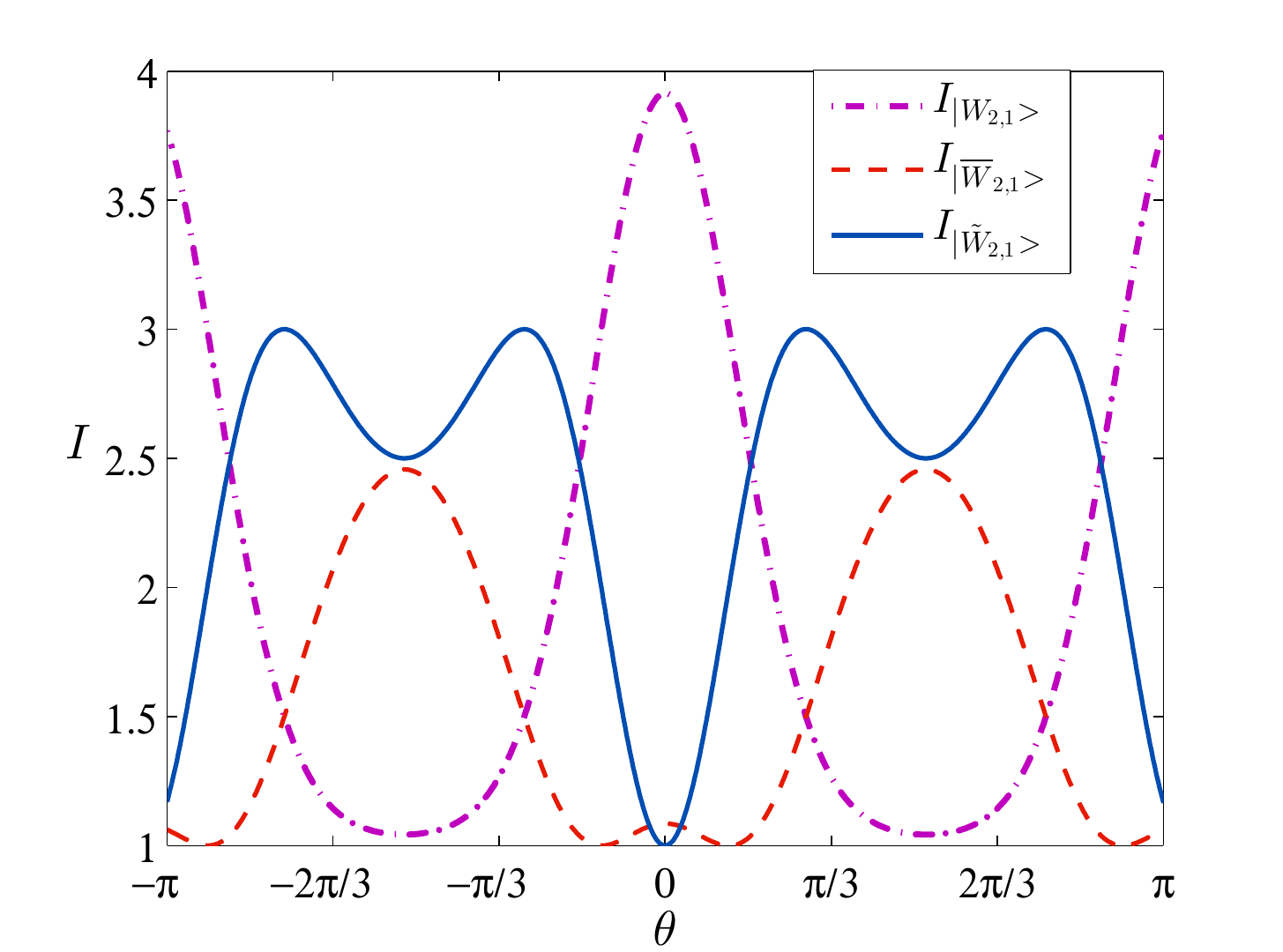} 
\label{fig:Line 2a}
}
\subfigure[]{
\includegraphics[width=7.5 cm,height=5.0 cm]{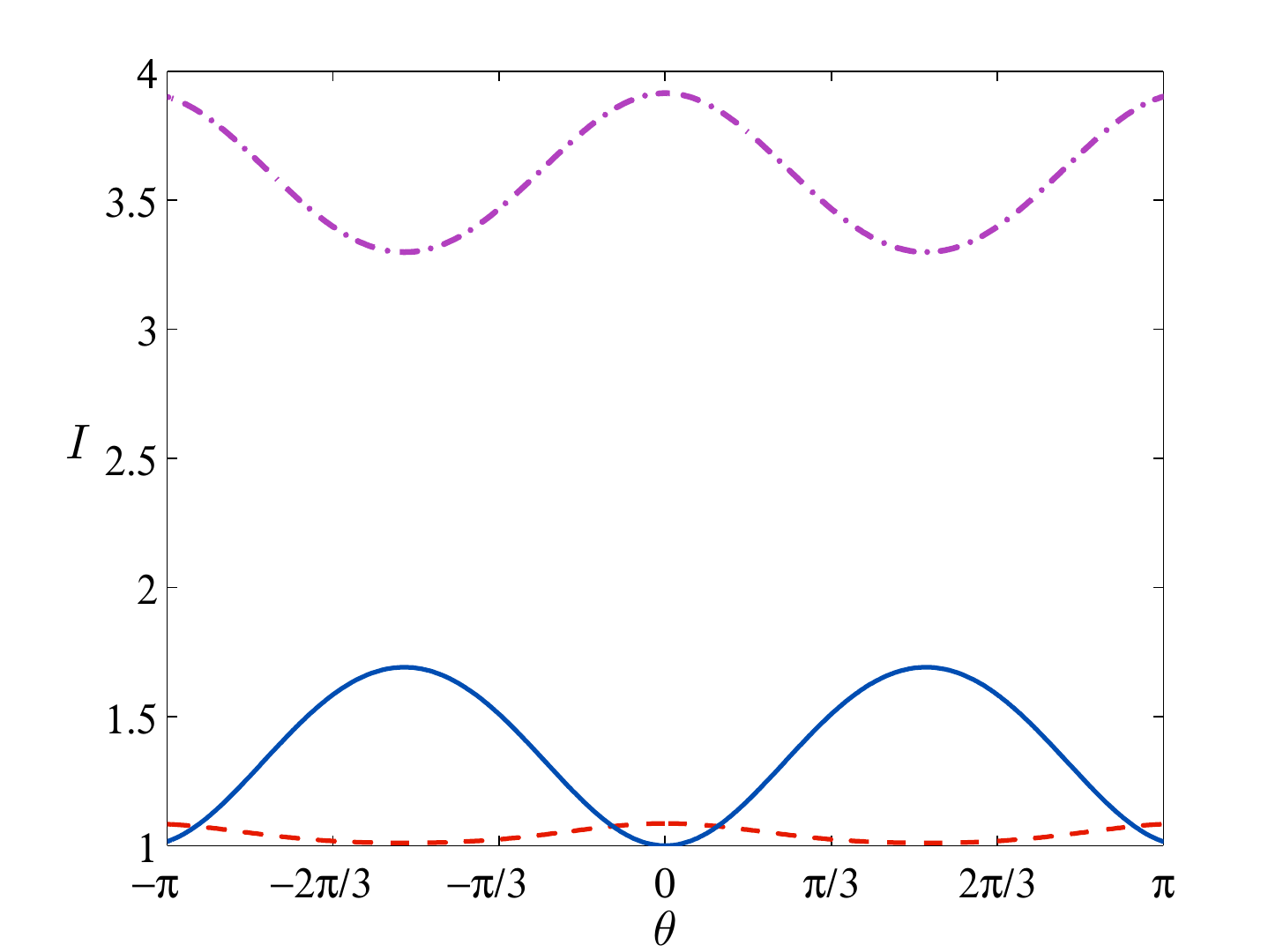}
\label{fig:Line 2b}
}
\caption{Line configuration: Intensity of the initial state $ | W_{2,1}\rangle  $(dot-dashed line), anti-symmetric state $| \overline{W}_{2,1}\rangle $(dashed line) and GHZ state $ | {\tilde{W}}_{2,1}\rangle $(solid line) as a function of the observation angle $\theta $ for \subref{fig:Line 2a} weak dipole-dipole interaction strength ($ \Omega_{12}=\Omega_{23}=0.29 \gamma $, for $ d =\frac{\lambda}{3} $) and \subref{fig:Line 2b} strong dipole-dipole interaction strength ($ \Omega_{12}=\Omega_{23}= 2.6 \gamma $, for $ d =\frac{\lambda}{10} $ ).}
\label{fig:Line 1}
\end{figure}

It has been observed earlier, the presence of dipole coupling between the atoms causes a mixing of the energy levels and creates different entangled W states.  The one-atom excited states give rise to one class of W states while the two-atom excited states give rise to a second class of W states.  One of the possible W state for the two-atom excitation is given by,
\begin{equation}\label{eq:lineW21}
| W_{2,1}\rangle =\frac{1}{2}\left[ |110 \rangle +|011 \rangle +\sqrt{2}|101 \rangle \right],~~\lambda_{1}=\sqrt{2}g+\dfrac{\omega}{2}
\end{equation}
where $ \lambda_{1} $ is the eigenvalue. The corresponding intensity pattern can be exactly calculated,
{\begin{footnotesize}\begin{equation}
I_{| W_{2,1}\rangle} =2+\frac{1}{2}\left[ \cos(\phi_{1}-\phi_{3})+\sqrt{2}\{\cos(\phi_{1}-\phi_{2})+\cos \left(\phi_{2}-\phi_{3}\right)\} \right]
\end{equation}\end{footnotesize}}
The intensity $ I_{| W_{2,1}\rangle} $  shows an angular dependence and exhibits a maximum value, as seen from Fig \ref{fig:Line 1},
\begin{equation}
[I_{| W_{2,1}\rangle}]^{Max}=3.914, \nonumber
\end{equation}
for $ \theta =0, \pi $.  The maximum intensity is higher than the corresponding intensity of the separable state, arising from the non-zero quantum correlations in case of W-state.  Comparing with the result of \cite{r17, r19, r20, r21}, the presence of dipole coupling seems to have little effect on the maxima and minima of the intensity for this W-state.  However, the behaviour in the regions between the maxima and minima has significant difference in the presence and absence of the interaction, as is evident from Fig \ref{fig:Line 1}.

Another type of anti-symmetric W state for the two atom in the excited state is obtained as,
\begin{equation}\label{eq:lineW21bar}
| \overline{W}_{2,1}\rangle =\frac{1}{2}\left[ |110 \rangle +|011 \rangle -\sqrt{2}|101 \rangle \right],~~\lambda_{2}=\dfrac{\omega}{2}-\sqrt{2}g,
\end{equation}
 and the corresponding intensity is given by
{\begin{footnotesize}\begin{equation}
I_{| \overline{W}_{2,1}\rangle} =2+\frac{1}{2}\left[ \cos(\phi_{1}-\phi_{3})-\sqrt{2}\{\cos(\phi_{1}-\phi_{2})+\cos \left(\phi_{2}-\phi_{3}\right)\} \right] 
\end{equation}\end{footnotesize}}
The intensity minima occurs at
\begin{equation}
[ I_{| \overline{W}_{2,1}\rangle}]^{Min}=1.086, \nonumber
\end{equation}
when $ \theta =0, \pi $. 

A second type of anti - symmetric state, of the GHZ type, with two atoms in excited state,
\begin{equation}
| {\tilde{W}}_{2,1}\rangle =\frac{1}{\sqrt{2}}\left[ |011 \rangle -|110 \rangle \right],~~\lambda_{3}=\dfrac{\omega}{2}
\end{equation}
 has the intensity form 
\begin{equation}
|I_{|{\tilde{W}}_{2,1}\rangle} =2- \cos(\phi_{1}-\phi_{3}).
\end{equation}
The angular intensity distribution of different W states for two different dipole-dipole coupling strengths is shown in figure \ref{fig:Line 1}.
\begin{figure}[ht!]
\centerline{\includegraphics[width=7.5 cm,height=6.0 cm]{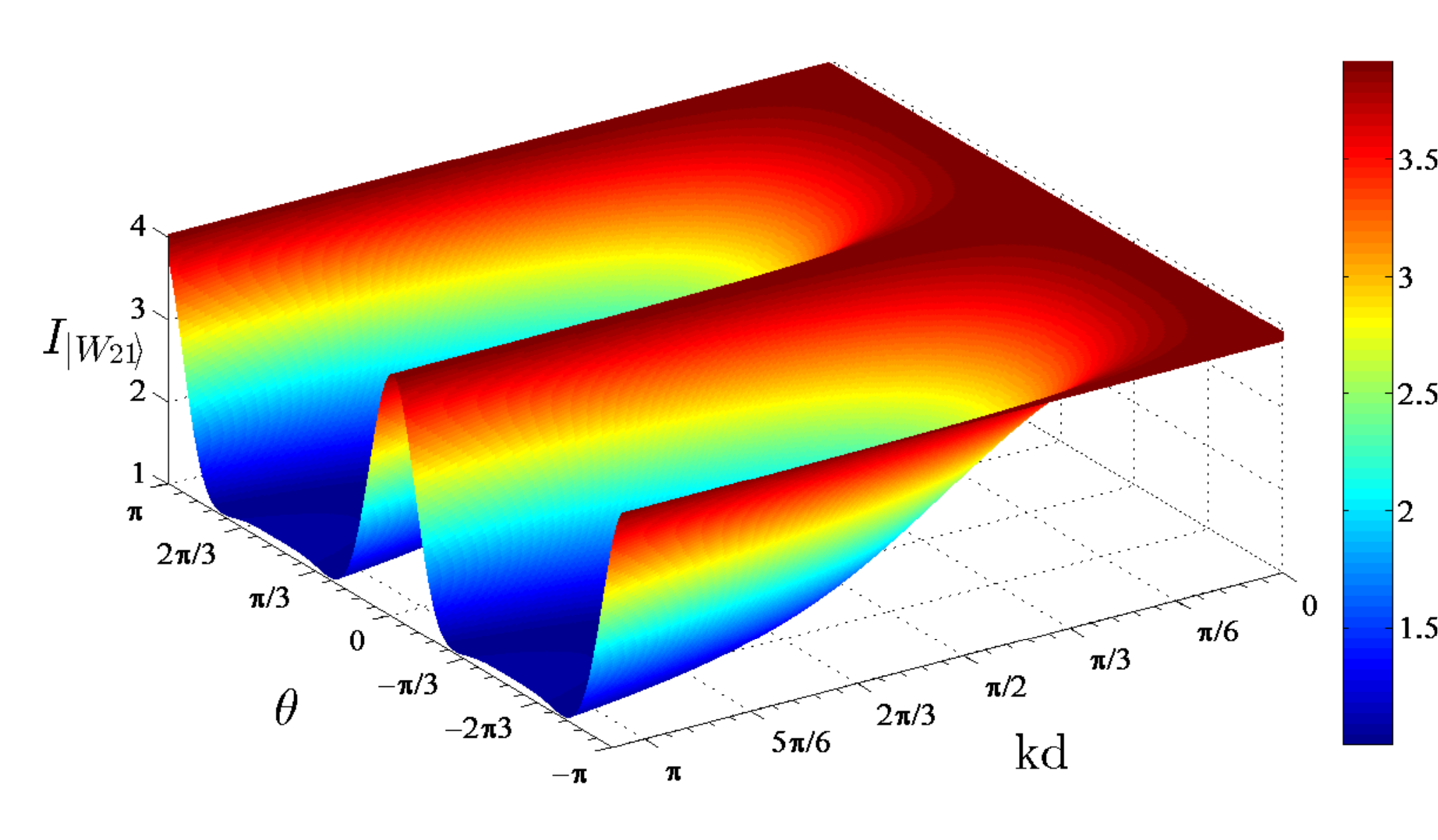}}
\caption[]{(Color online) Surface Plot of Intensity $I_{| W_{2,1}\rangle} $ as a function of observation angle $\theta$ and interatomic distance $kd$}
\label{fig:surf 1}
\end{figure}
\begin{figure}[ht!]
\centerline{\includegraphics[width=7.5 cm,height=6.0 cm]{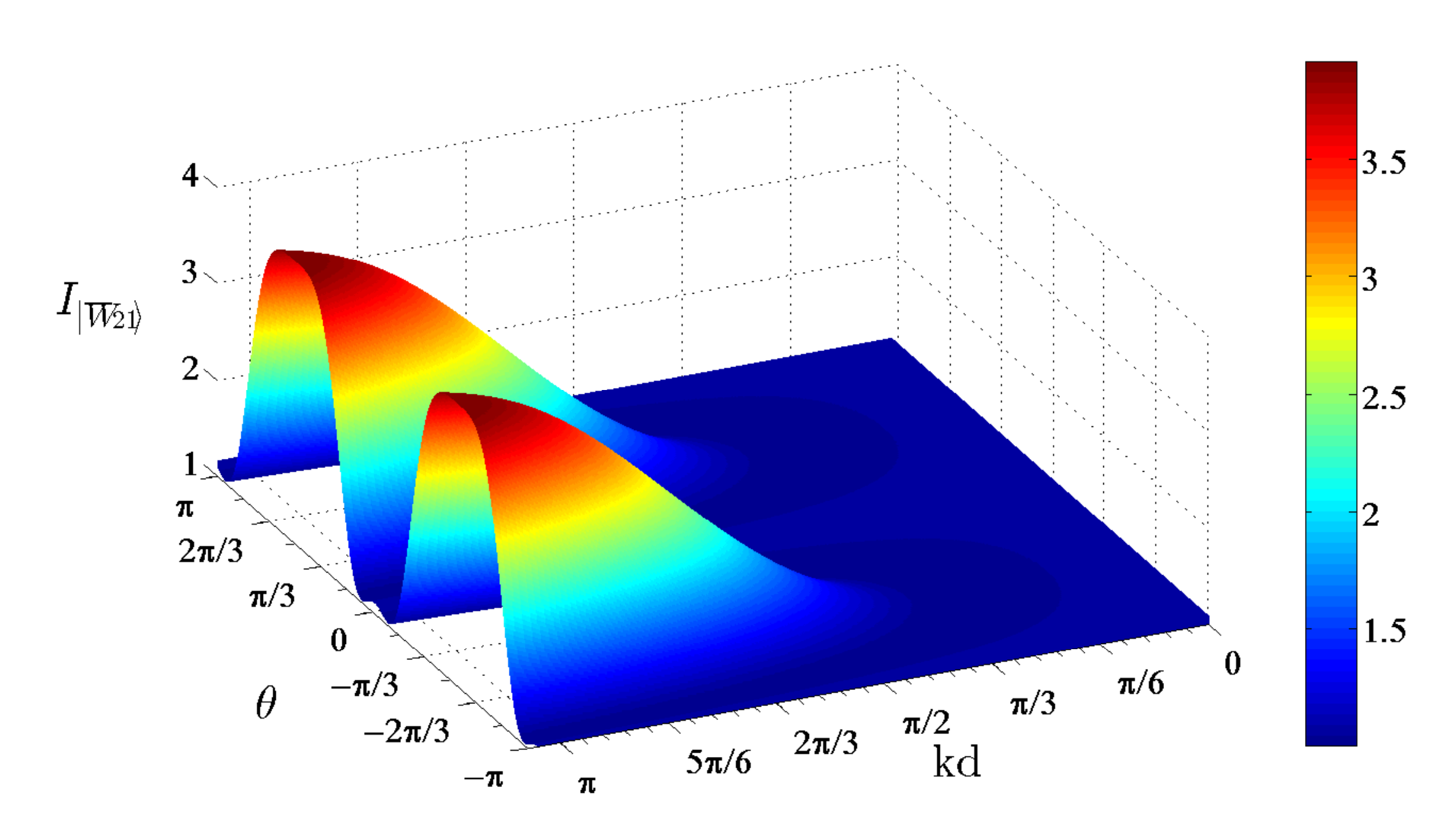}}
\caption[]{(Color online) Surface Plot of Intensity $I_{| \overline{W}_{2,1}\rangle}$ as a function of observation angle $\theta$ and interatomic distance $kd$}
\label{fig:surf 2}
\end{figure}
When the atoms are arranged in a line, the intensity from each member of the states ($ |W_{21}>,| \overline{W}_{2,1}\rangle $ and $| {\tilde{W}}_{2,1}\rangle $) is markedly different from one other.  The intensity from the  anti-symmetric  $ | \overline{W}_{2,1}\rangle$ state and the GHZ - state $| {\tilde{W}}_{2,1}\rangle $ is showing a clear complementary behaviour at certain observation angles, for example at $ \theta = 0, \pm \pi \ /2$.  The results here, in which dipole-dipole interaction is included [eq.(\ref{eq:lineW21}), eq.(\ref{eq:lineW21bar})] are showing different behaviour from that of \cite{r17}. 

To gain a clearer visual understanding of this feature, three-dimensional plots of the intensities for both the  state $| W_{2,1}\rangle$ as well as the anti - symmetric state $ | \overline{W}_{2,1}\rangle$ are presented in figures \ref{fig:surf 1} and \ref{fig:surf 2} for a range of inter atomic distances and as a function of the observation angle.  The three - dimensional view clearly shows the separated sub-radiant (blue) and super-radiant (red) regions.  We also observe that, as the inter atomic distance is decreased, the super radiant behaviour becomes more pronounced for the W-state $| W_{2,1}\rangle$, where as, the sub radiant behaviour becomes stronger for the anti - symmetric  W-state  $ | \overline{W}_{2,1}\rangle$.  In addition, we can clearly see the complementary behaviour of the intensity for  these two W - states. 

\subsection{Photon Statistics}
In addition to the intensities, study of photon statistics, inferred from the second order intensity - intensity correlation function, enables one to gain insight into the characteristics of the emitted radiation field.  In this section, such a study for the emitted photons is carried out.  In particular, we concentrate on the super / sub-radiant behaviour of the emitted radiation. 
\begin{figure}[ht!]
\centerline{\includegraphics[width=9 cm,height=6.5 cm]{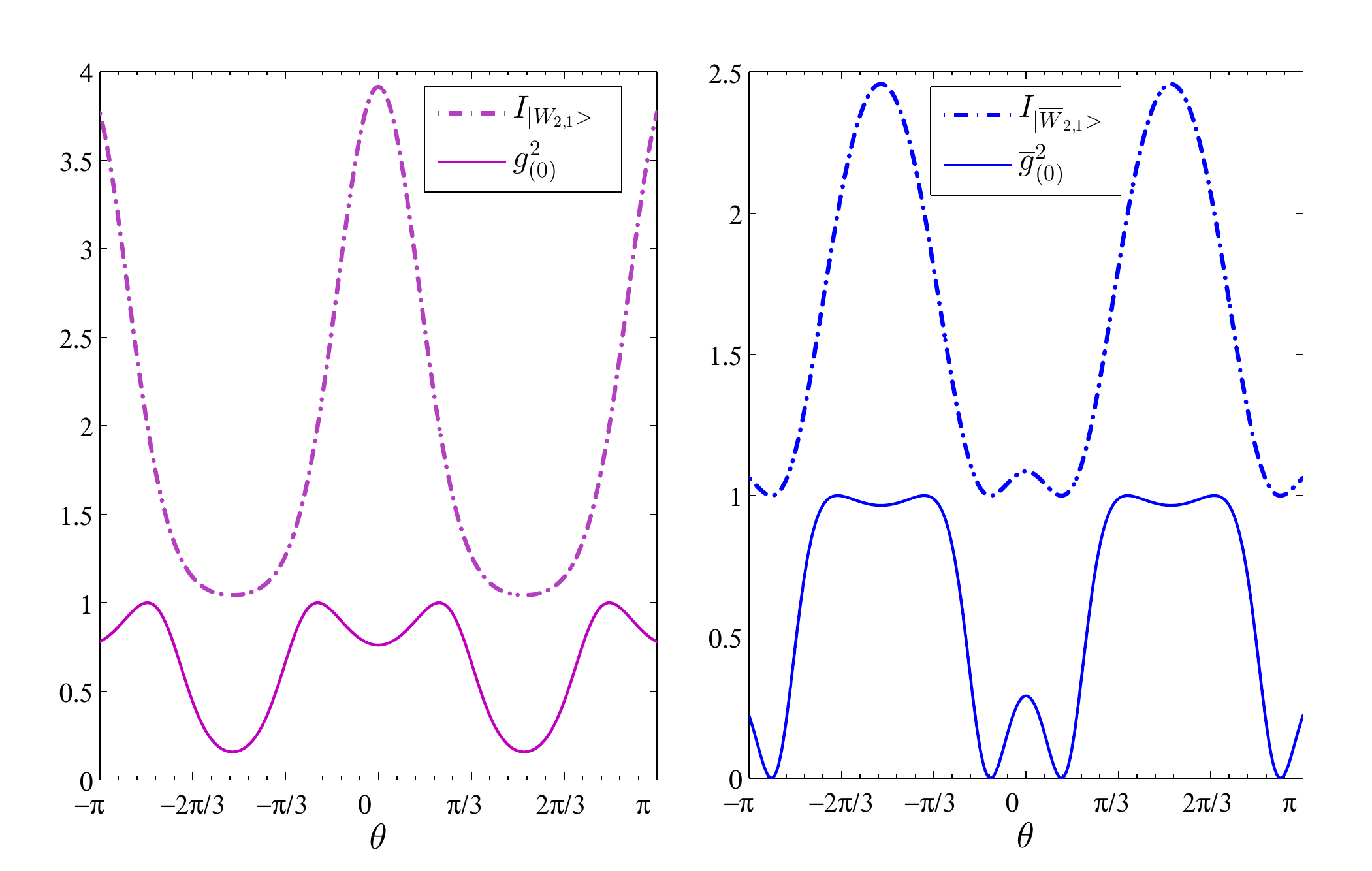}}
\caption[]{(Color online)Line configuration: The intensity (dot-dashed line) and the second order correlation function of radiation field (solid line), emitted by two atoms which are initially in (a) the  $| W_{2,1}\rangle $ state and (b)$| \overline{W}_{2,1}\rangle $ as a function of observation angle $ \theta $ for $ kd=\dfrac{2\pi}{10} $.}
\label{fig:line correlation}
\end{figure}
The second-order (intensity-intensity) correlation function of the radiation field, with zero time lag, is defined as
\begin{equation}
g^{(2)}(0)=\dfrac{\langle E^{-}E^{-}E^{+}E^{+}\rangle}{\langle E^{-}E^{+}\rangle \langle E^{-}E^{+}\rangle}
\end{equation}
Violation of classical inequalities of this second-order correlation function points to non-classical character. 
In particular, $ g^{(2)}(0) < 1 $, a non-classical signature, refers to the sub-Poissonian photon statistics of the radiation field, while  $ g^{(2)}(0) > 1 $ corresponds to the super-Poissonian.  The reference value of  $ g^{(2)}(0) = 1 $ is a characteristic of  the Poissonian photon statistics, as exhibited by the laser field.  After a lengthy calculation, the second-order correlation function of the radiation field, emitted by three atoms, which are initially in the  $ |W_{21}> $ state is obtained as 
{\begin{scriptsize}\begin{equation}
g^{(2)}(0)=\\
\dfrac{4+2[\cos(\phi_3 -\phi_1)+\sqrt{2}(\cos(\phi_1 -\phi_2)+\cos(\phi_2 -\phi_3))]}{\left[2+\dfrac{1}{2}\{\cos(\phi_3 -\phi_1)+\sqrt{2}[\cos(\phi_1 -\phi_2)+\cos(\phi_2 -\phi_3)]\}\right]^{2}}
\end{equation}\end{scriptsize}}
The corresponding expression for the intensity correlation, when the two atoms are  initially in $| \overline{W}_{2,1}\rangle $, denoted here by $\bar{g}^{(2)}(0)$, is given as 
{\begin{scriptsize}\begin{equation}
\bar{g}^{(2)}(0)=\\
\dfrac{4+2[\cos(\phi_3 -\phi_1)-\sqrt{2}(\cos(\phi_1 -\phi_2)+\cos(\phi_2 -\phi_3))]}{\left[2+\dfrac{1}{2}\{\cos(\phi_3 -\phi_1)-\sqrt{2}(\cos(\phi_1 -\phi_2)+\cos(\phi_2 -\phi_3))\}\right]^{2}}
\end{equation}\end{scriptsize}}

In figure \ref{fig:line correlation}, the results corresponding to the intensity as well as the second order correlation function are presented for the   $| W_{2,1}\rangle$ state as well as the  anti - symmetric $ | \overline{W}_{2,1}\rangle$  state for an inter-atomic distance $ kd=\dfrac{2\pi}{10} $.  
It is observed that for both the $| W_{2,1}\rangle$ as well as the anti - symmetric $ | \overline{W}_{2,1}\rangle$ states, the correlation function  is exhibiting non classical nature throughout most of the region, except at four different observation angles  in each case (as can be clearly seen from the figure), at which it is attaining the value of 1 corresponding to the poissonian character. 
For the range of observation angles corresponding to relatively smaller values of the correlation functions, which points to stronger non classical character, the corresponding intensity is showing sub radiant behaviour.  Whereas for the range of observation angles for relatively larger values of the correlation function, which can be interpreted as weak non-classicality, the intensity is super-radiant.  Similar association between the strong/weak non classicality and sub/super radiant intensity is seen for the anti-symmetric $ | \overline{W}_{2,1}\rangle$ state.  In figure \ref{fig:line correlation}   a careful examination of  (a)  and (b) shows that wherever the intensity of one of the W - states is subradiant, that of the other state is superradiant and vice-versa.  In summary, both the W- states are showing non-classicality mostly, with their intensities showing complementary behaviour.  


\section{The intensity characteristics of light emitted by three atoms in a Loop-configuration}
 \begin{figure}[h!]
 \centering
\includegraphics[width=7.0 cm,height=6.0 cm]{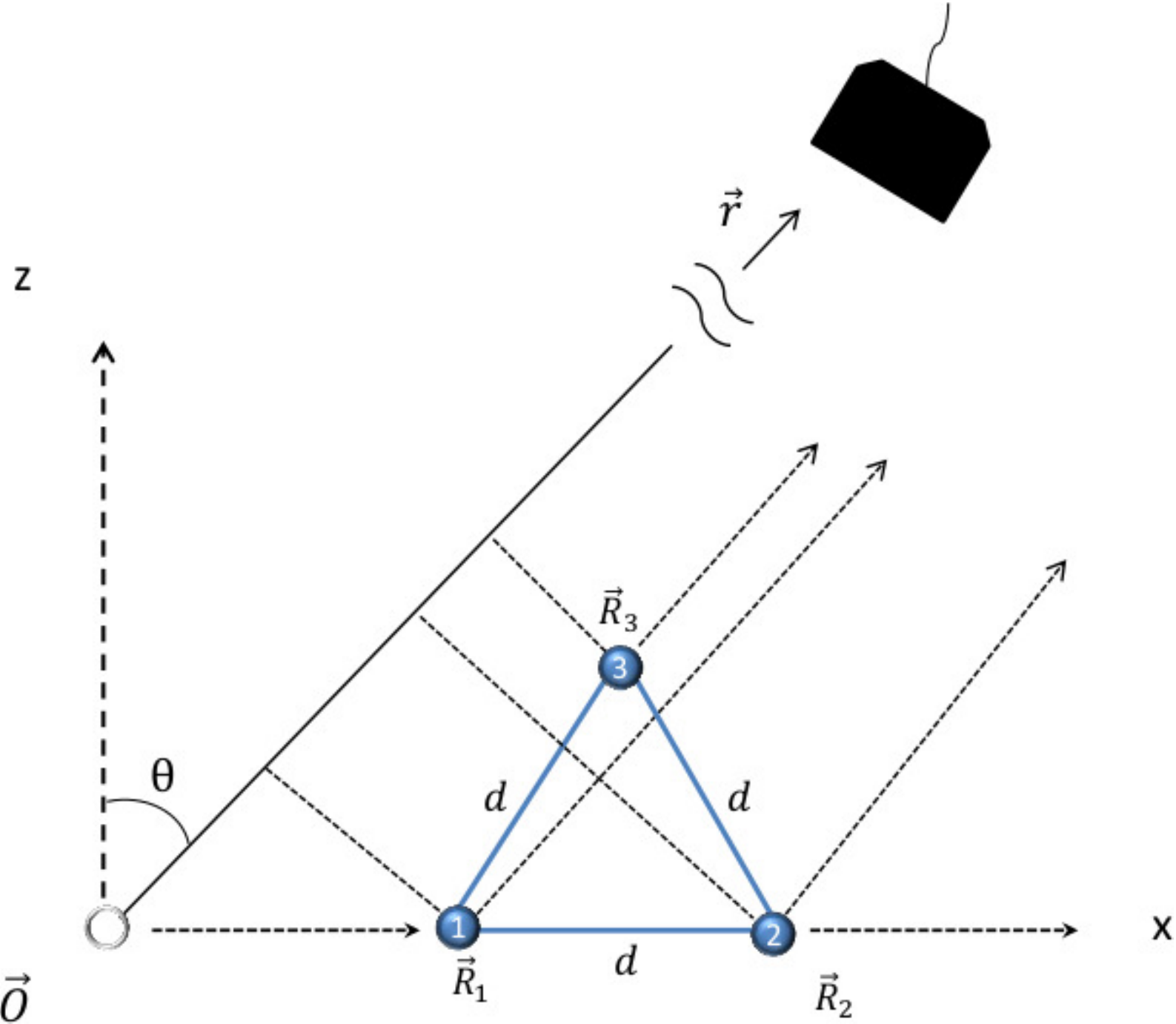}
\caption[]{(Color online) Schematic diagram of the system: where three identical and equidistant two-level atoms are localized at the vertices of an equilateral triangle.  A detector at position $ \bar{r} $ records the photon scattered by the atoms, in the far field regime.}
\label{fig:Three}
\end{figure}
The far - field intensity from different GHZ - states and the symmetric W - state  $| W_{2,1}\rangle$ corresponding to the loop configuration shows entirely different  behaviour from that of the line-configuration discussed in the previous section.  In the loop configuration, the schematic of which is shown in Fig. \ref{fig:Three}, the relative optical phase accumulated by a photon emitted at $ \vec{R}_{j} $ and detected at $ \vec{r} $ is given by 
\begin{equation}
\phi_{1}(\vec{r})\equiv \phi_{1}=k\vec{n}.\vec{R}_{1}=kd\sin{\theta}
\end{equation} 
\begin{equation}
\phi_{2}(\vec{r})\equiv \phi_{2}=k\vec{n}.\vec{R}_{2}=2kd\sin{\theta}
\end{equation} 
\begin{equation}
\phi_{3}(\vec{r})\equiv \phi_{3}=k\vec{n}.\vec{R}_{3}=\frac{3kd\sin{\theta}+\sqrt{3}kd\cos{\theta}}{2}
\end{equation} 
 \begin{figure}[ht!]
\centering
\subfigure[]{
\includegraphics[width=7.5 cm,height=5.0 cm]{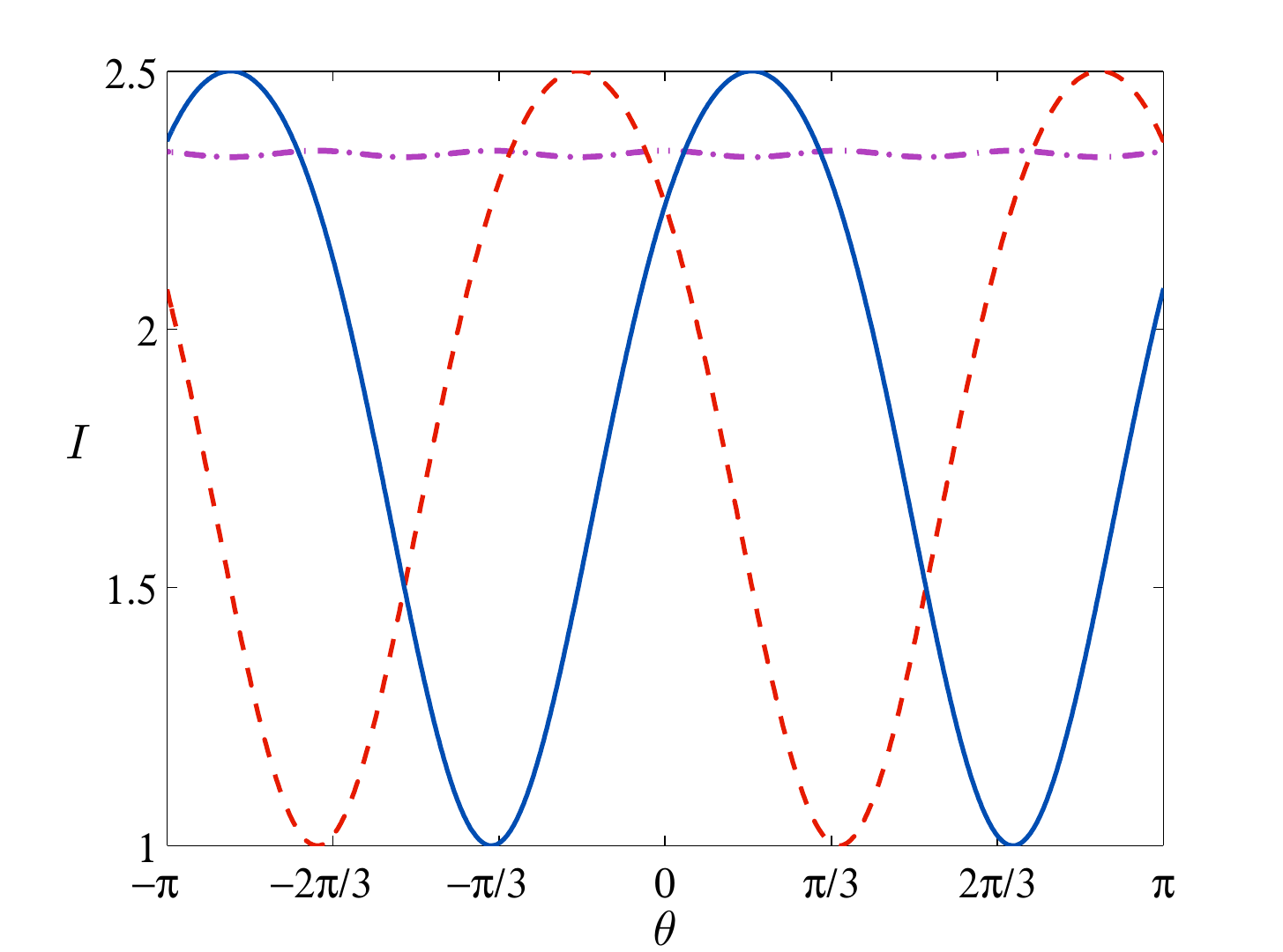} 
\label{fig:Loop 2a}
}
\subfigure[]{
\includegraphics[width=7.5 cm,height=5.0 cm]{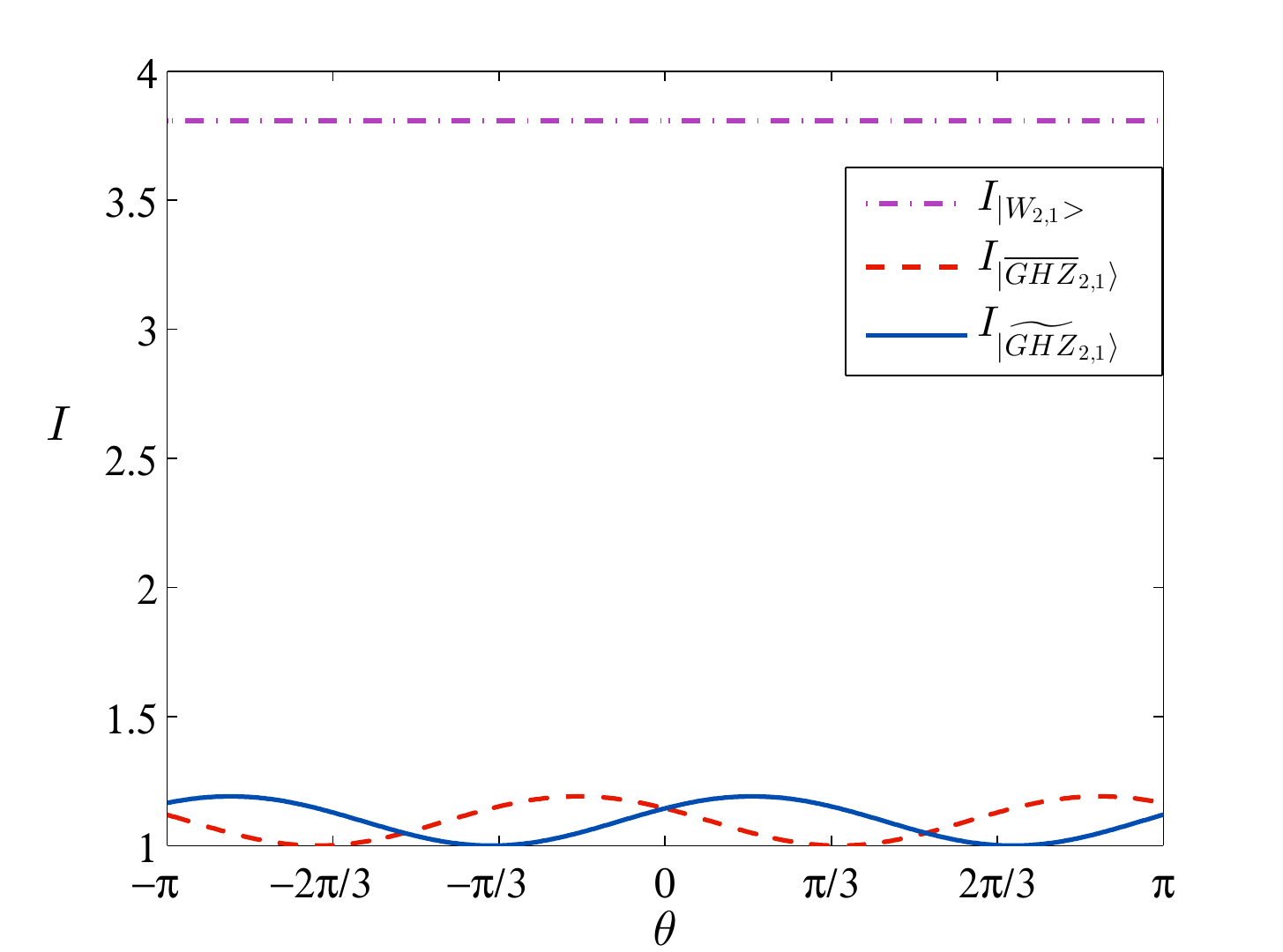}
\label{fig:Loop 2b}
}
\caption{Loop configuration: Intensity of the initial symmetric state $ | W_{2,1}\rangle  $(dot-dashed line), GHZ state$| \overline{GHZ}_{2,1}\rangle $ (dashed line)  and second type GHZ state $|{\widetilde{GHZ}}_{2,1}\rangle$ (solid line)  as a function of the observation angle $\theta $ for the inter atomic distance \subref{fig:Loop 2a} $ d =\frac{\lambda}{3} $(corresponding $ \Omega_{12}=\Omega_{23}=0.29 \gamma $) and \subref{fig:Loop 2b} $ d =\frac{\lambda}{10} $ ($ \Omega_{12}=\Omega_{23}= 2.6 \gamma $).}
\label{fig:Loop 2}
\end{figure}
\begin{figure}[ht!]
\centerline{\includegraphics[width=7.0 cm,height=7.0 cm]{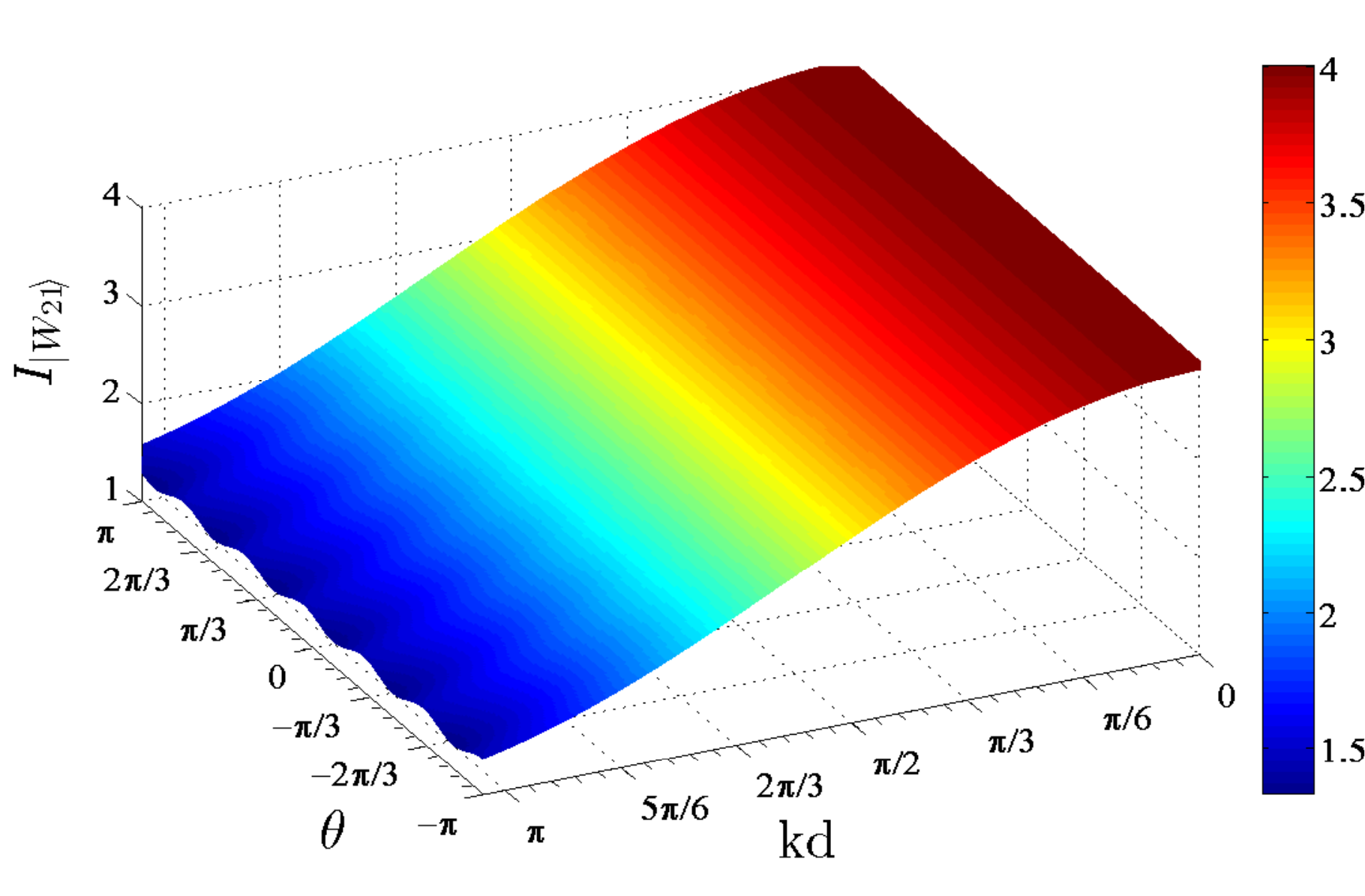}}
\caption[]{(Color online) surface plot of $I_{| W_{2,1}\rangle}$ as a function of $\theta $ and $ kd $}
\label{fig:surf loop}
\end{figure}

 For this configuration, the resulting  symmetric W - state given by
\begin{equation}\label{Eq:loopW21}
| W_{2,1}\rangle =\frac{1}{\sqrt{3}}\left[ |110 \rangle +|011 \rangle +|101 \rangle \right],~~\lambda_{4}=2g+\dfrac{\omega}{2} 
\end{equation}
and the corresponding  intensity is found to be
{\begin{footnotesize}\begin{equation}
I_{| W_{2,1}\rangle} =2+\frac{2}{3}\left[ \cos(\phi_{1}-\phi_{3})+\cos(\phi_{1}-\phi_{2})+\cos(\phi_{2}-\phi_{3}) \right].
\end{equation}\end{footnotesize}}

 Similarly, one can construct two types of GHZ states when two atoms are present in the excited state.
\begin{equation}
| \overline{GHZ}_{2,1}\rangle =\frac{1}{\sqrt{2}}\left[ |101 \rangle -|110 \rangle \right],~~\lambda_{5}=\dfrac{\omega}{2}-g,
\end{equation}
 with the intensity,
\begin{equation}
I_{| \overline{GHZ}_{2,1}\rangle} =2-\cos(\phi_{2}-\phi_{3})  ,
\end{equation}
\begin{figure}[h!]
\includegraphics[width=9.0 cm]{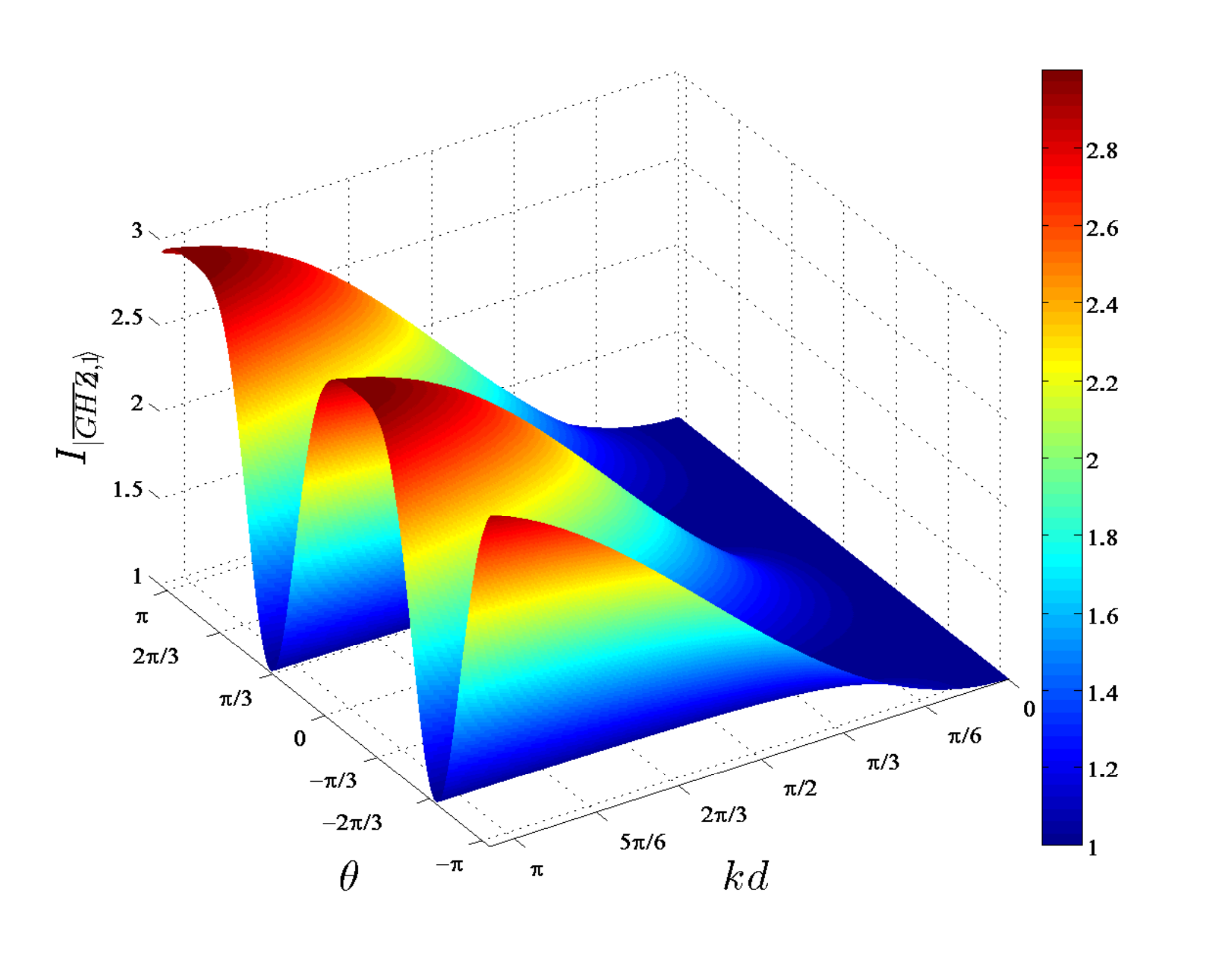}
\caption[]{(Color online) surface plot of $I_{|\overline{GHZ}_{2,1}\rangle}$}
\label{fig:surf loop2}
\end{figure}

For the second GHZ case,
\begin{equation}
|{\widetilde{GHZ}}_{2,1}\rangle =\frac{1}{\sqrt{2}}\left[ |011\rangle -|110\rangle \right],~~\lambda_{6}=\dfrac{\omega}{2}-g,
\end{equation}
the intensity of which  is given by
\begin{equation}
I_{| {\widetilde{GHZ}}_{2,1}\rangle} =2-\cos(\phi_{1}-\phi_{3}) 
\end{equation}
The intensity profile of different GHZ - states and the symmetric W - state  $| W_{2,1}\rangle$  as defined above, for the loop configuration, is shown in Fig.\ref{fig:Loop 2} for two different values of the dipole coupling strength.  The intensity in case of the symmetric W - state is  super-radiant for the chosen dipole coupling strength and nearly independent of the observation angle, meaning thereby that the emitted radiation is nearly isotropic. 	Another interesting feature about the behaviour of the radiation from this W- state is that for stronger dipole coupling strength which results in stronger superradiant character, the small fluctuations about a base-line value,  in the emitted radiation disappear  and the emitted radiation is indeed isotropic.  As for the two GHZ - states that we have defined for the case of two atoms initially in the excited states, the intensity is periodically varying from subradiant to superradiant values as the observation angle is changed for a relatively larger interatomic distance ($ kd = \frac{2 \pi}{3}$). For a smaller interatomic distance ($ kd = \frac{2 \pi}{10}$), the intensity remains subradiant for all observation angles.  In addition, the radiation from these two states is out of phase, the phase difference being a function of the interatomic distance,  as can clearly be seen from parts (a) and (b) of Fig.\ref{fig:Loop 2}. Three dimensional plots of the radiated intensity of the symmetric W - state and the GHZ - states  of Fig.\ref{fig:Loop 2}  are  shown  in Fig.\ref{fig:surf loop} and Fig.\ref{fig:surf loop2} respectively.  In case of the W - state (\ref{fig:surf loop}), one can clearly see how the intensity goes from subradiant to superradiant with gradual decrease in the interatomic distance. The vanishing of the flutuations of the intensity is also clearly visible. Comparing the figures  \ref{fig:surf loop} and \ref{fig:surf loop2}, one can see the sensitive dependence of the intensity pattern of the GHZ - state 
 on both the observation angle as well as the interatomic distance and near-absence of the same for the W - state.   One observes that the intensity can be tuned by  changing the interatomic distances which in turn result in change in the dipole coupling strengths which suggests the exciting possibility of optical probing of the entanglement characteristics. 
  
  \subsection{Photon statistics} 
In this section, numerical results of the photon statistics of the radiation field,  when two of the three atoms in loop configuration are in the excited state, are presented .
 The second-order equal - time correlation function  when the system is initially in the $ |W_{21}> $ state is given by,
{\begin{scriptsize}\begin{equation}
g^{(2)}(0)=\dfrac{4+\frac{8}{3}\left[\cos(\phi_1 -\phi_2)+\cos(\phi_2 -\phi_3)+\cos(\phi_3 -\phi_1)\right]}{\left[2+\dfrac{2}{3}\{\cos(\phi_1 -\phi_2)+\cos(\phi_2 -\phi_3)+\cos(\phi_3 -\phi_1)\}\right]^{2}}.
\end{equation}\end{scriptsize}}

\begin{figure}[ht!]
\centerline{\includegraphics[width=9.5 cm,height=5.0 cm]{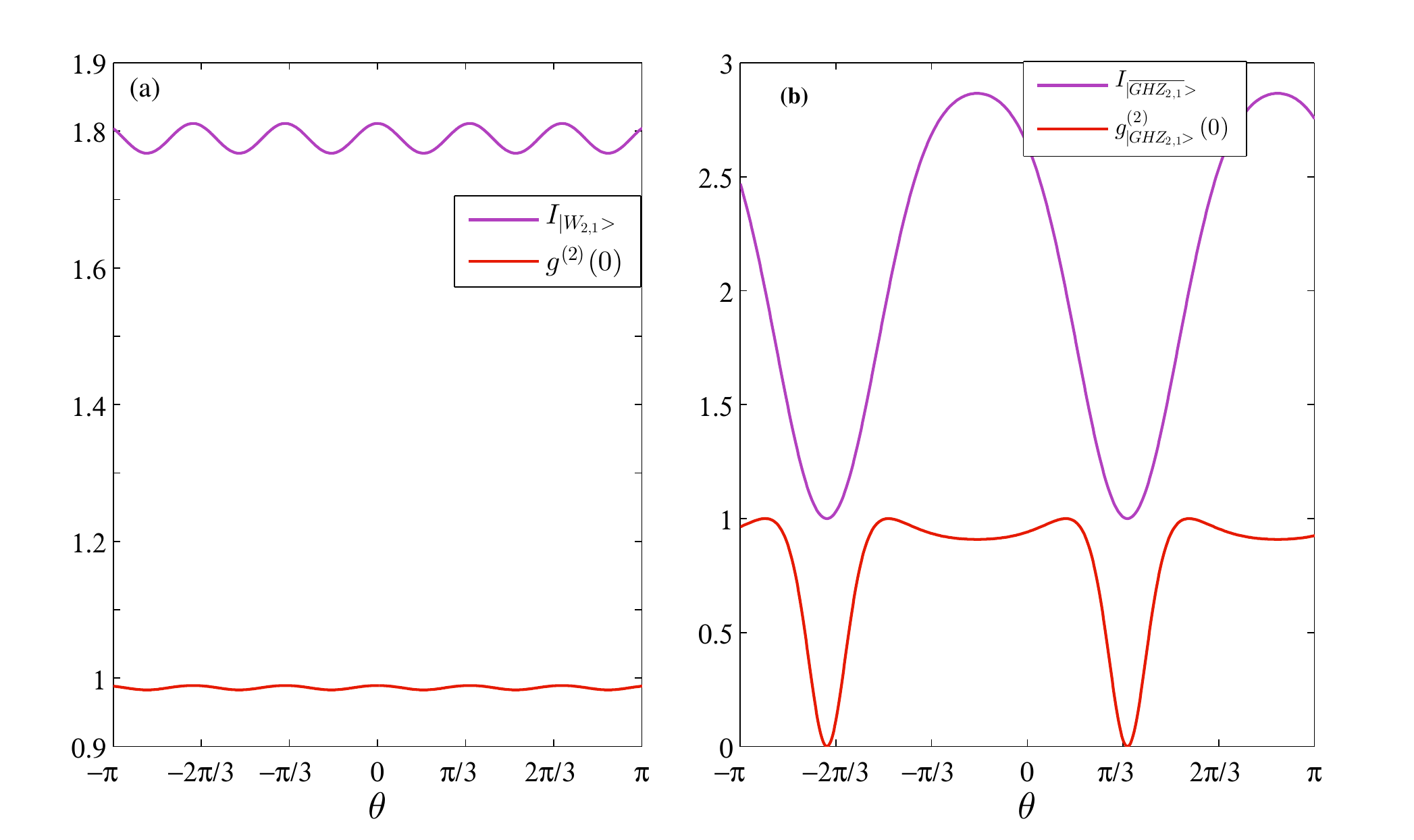}}
\caption[]{(Color online)  Loop configuration: The intensity (magenta line) and the second order correlation function (red line) of radiation field for (a) the symmetric $ |W_{21}> $ state and (b) $|\overline{GHZ}_{2,1}\rangle $ state, as a function of observation angle $ \theta $ for $ kd=\dfrac{5\pi}{6} $.}
\label{fig:loop correlation 1}
\end{figure}

\begin{figure}[ht!]
\centerline{\includegraphics[width=10 cm,height=5.0 cm]{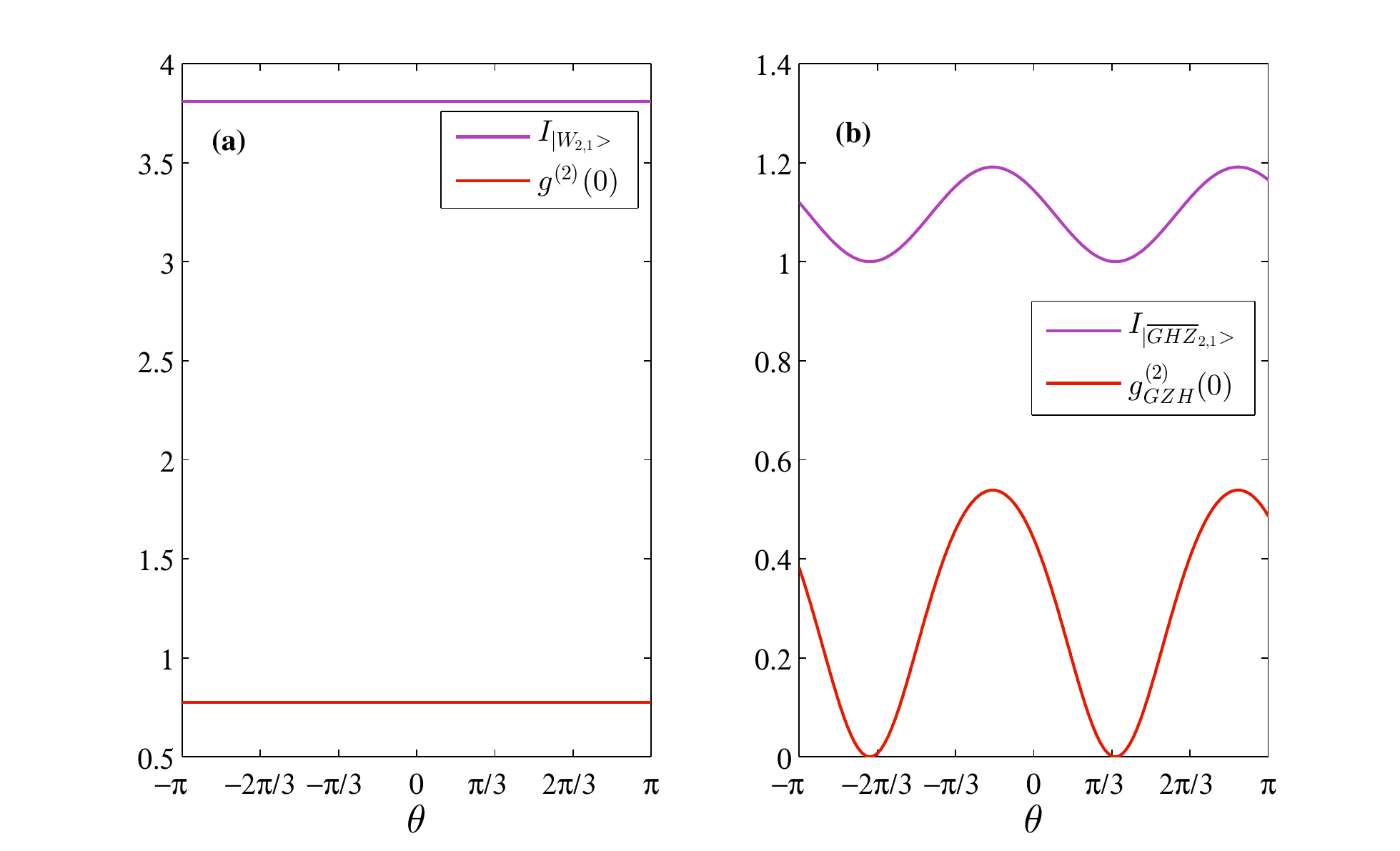}}
\caption[]{(Color online)  Loop configuration: The intensity (magenta line) and the second order correlation function (red line) of radiation field for (a) the symmetric $ |W_{21}> $ state and (b) $|\overline{GHZ}_{2,1}\rangle $ state, as a function of observation angle $ \theta $ for $ kd=\dfrac{2\pi}{10} $.}
\label{fig:loop correlation 2}
\end{figure}

In case of the symmetric  $ |W_{21}> $ state, a careful observation of Fig.\ref{fig:loop correlation 1}(a) and Fig.\ref{fig:loop correlation 2} (a) shows that the photon statistics remains subpoissonian for both the weak and strong coupling strengths, whereas the intensity changes from subradiant to superradiant.   In the GHZ - state ( Fig.\ref{fig:loop correlation 1}(b) and Fig.\ref{fig:loop correlation 2} (b)), it is however quite different in the sense that for  weak coupling strengths, the radiation emitted is showing periodic variations between subradiant and superradiant regimes and with increased coupling strength these oscillations are confined to subradiant regime alone.  The corresponding photon statistics is similar to that of the W - state, viz., the nonclassicality as seen from the behaviour of correlation function is becoming stronger with increase in coupling strength.  

In summary, for the case of GHZ - state, the photon statistics is varying from sub-poissonian to poissonian with clear dependence on the angle of observation whereas in the case of $ |W_{21}> $ state it is  remaining subpoissonian throughout and is essentially independent of the observation angle.  Hence,  the entanglement property of the three atom dipolar coupled system is found to yield characteristically distinct far-field radiation pattern, as also the photon statistics, in different configurations.   It is observed, from the current study that, this behaviour can be tuned  by appropriate change of the atomic distances as also the dipolar couplings.  
  
 \section{CONCLUSION}
 
 In conclusion,  the fact that entanglement arises due to superposition of states, with the degree of entanglement being controlled by the nature of superposition, offers the possibility of its signature on the radiation pattern in the present case of coupled dipolar system.  It is interesting to note that distinct entangled configurations, in a tripartite system studied here,  indeed revealed distinct radiation pattern as also intensity - intensity correlation. This opens a way for optical probing of the entanglement characteristics.  
 
\section{ACKNOWLEDGEMENTS}
Shaik Ahmed acknowledges the financial support from the University Grants Commission, India.

\end{document}